\documentclass[useAMS,usenatbib]{mn2e}
\usepackage{graphicx} 
\usepackage{times} 

\def\kms{km ${\rm s}^{-1}$}

\def\ch2{$\chi^2$}
\def\dg{$^{\circ}$}

\def\Lo{L$_\odot$}
\def\Mo{M$_\odot$}

\def\kms {\hbox{${\rm km\ s}^{-1}$}}

\def\scm  {$\hbox{{\rm cm}}^{-2}$}    
\def\arcsec {\hbox{$^{\prime\prime}$}}

\def \AL {$\alpha $}     
\def \HI {H{\sc \,i}}

\def\lapp{\ifmmode\stackrel{<}{_{\sim}}\else$\stackrel{<}{_{\sim}}$\fi}
\def\gapp{\ifmmode\stackrel{>}{_{\sim}}\else$\stackrel{>}{_{\sim}}$\fi}
\def\bsp_small{\vspace{0.5cm}\small\noindent This paper has been typeset
from a \TeX/\LaTeX\ file prepared by the author.\normalsize}

\title[21-cm absorption at $z_{\rm abs}=0.656$ towards 3C\,336]{Detection of broad 21-cm absorption at {\boldmath$z_{\rm abs}=0.656$} in the complex sight-line towards 3C\,336}
\author[S. J. Curran et al.]{S. J. Curran$^{1}$\thanks{E-mail: sjc@phys.unsw.edu.au}, P. Tzanavaris$^{2}$, M. T. Murphy$^{3}$, J. K. Webb$^{1}$ and Y. M. Pihlstr\"{o}m$^{4}$\\ 
$^{1}$School of Physics, University of New South Wales, Sydney NSW 2052, Australia\\
$^{2}$Institute of Astronomy and Astrophysics, National Observatory of Athens, I.Metaxa \& V.Paulou
152 36 Penteli, Greece\\
$^{3}$Institute of Astronomy, Madingley Road, Cambridge CB3 0HA, UK\\
$^{4}$Department of Physics and Astronomy, The University of New Mexico, Albuquerque, NM 87131, USA}
\begin{document}

\date{Accepted ---. Received ---; in original form ---}

\pagerange{\pageref{firstpage}--\pageref{lastpage}} \pubyear{2007}

\maketitle

\label{firstpage}

\begin{abstract}
We report the detection of 21-cm absorption at $z_{\rm abs}=0.656$
towards 1622+238 (3C\,336). The line is very broad with a Full-Width
Half Maximum (FWHM) of 235 \kms, giving a velocity integrated optical
depth of $\int\tau dv = 2.2\pm0.2$ \kms. The centroid of
the line is offset from that of the known damped Lyman-$\alpha$
absorption (DLA) system by $50$ \kms, and {\em if} the Lyman-$\alpha$
and 21-cm absorption are due to the same gas, we derive a spin
temperature of $T_{\rm s}\leq60$ K, which would be the lowest yet in a
DLA. The wide profile, which is over four times wider than that of any
other DLA, supports the hypothesis that the hydrogen
absorption is occurring either in the disk of a large underluminous
spiral or a group of dim unidentified galaxies, associated with the
single object which has been optically identified at this redshift.

\end{abstract}


\section{Introduction}\label{intro}

Redshifted absorption systems lying along the sight-lines to distant
quasars are important probes of the early-to-present day Universe.  In
particular, damped Lyman-$\alpha$ absorption systems (DLAs), where
$N_{\rm HI}\ge2\times10^{20}$ \scm, are useful as they account for at
least 80\% of the neutral gas mass density in the Universe
\citep{phw05}. Since the Lyman-$\alpha$ transition occurs in the
ultra-violet band, direct ground based observations of this
transition are restricted to redshifts of $z\gapp1.7$, although
observations of the \HI ~spin-flip transitions at $\lambda_{\rm
rest}=21$ cm can probe neutral hydrogen from $z=0$, thereby providing a useful
complement to the high redshift optical data. Furthermore, 21-cm
observations can be used to confirm and complement reported variations
in the fundamental constants over the history of the Universe
\citep{mwf03}. Unfortunately, redshifted radio absorption systems are
currently very rare with only 50 \HI\ 21-cm absorption systems known
(summarised in \citealt{cwm+06}), 17 of which are DLAs, with only nine
of these having sufficiently high quality UV/optical and 21-cm data
\citep{tmw+06}.  We are therefore currently undertaking a survey of,
as yet unsearched, DLAs occulting radio-loud quasars in order to
increase the number of 21-cm absorbing systems and in this letter we
report our first detection, in the intriguing DLA towards 1622+238
(3C\,336).

\section{Observations and Data Reduction}\label{obs}

The observations were performed on 9--10 July 2006 with the Westerbork
Synthesis Radio Telescope (WSRT) as part of our survey 
(described in \citealt{ctp+07}). The front-end was the UHF-high
receiver tuned to 857.68 MHz (or $z_{\rm abs}=0.6561$,
\citealt{rt00}), backed with a band-width of 5 MHz over 2048 channels
(dual polarisation), giving a channel width of 0.85 \kms. The two
orthogonal polarisations (XX \& YY) were used in order to allow the
removal of any polarisation-dependent radio frequency interference
(RFI). After the flagging out of time-dependent RFI, 12.0 hours of
good data remained, although there was some RFI remaining on some
baselines, particularly in the XX polarisation. After further
flagging, 63 full and partial baseline pairs remained. Additionally,
frequencies below 856.6 MHz (channels above 1500) were removed due to
a steep 1 Jy dip in the flux.

\begin{figure}
\vspace{11.5cm}
\includegraphics{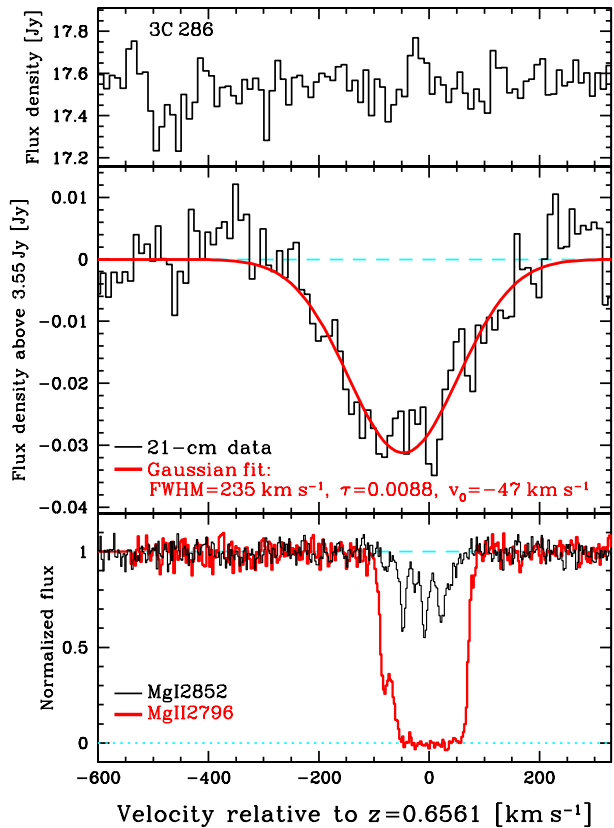}
\caption{Top: The bandpass calibrated spectrum of 3C\,286. Middle: The 21-cm spectrum of
1622+238. The Gaussian fit gives a peak line
depth of $30.8$ mJy (giving $\tau=0.0088$, against a 3.55 Jy continuum
flux density), a centroid at $-47\pm5$ \kms\ and a FWHM of 235 \kms\
for the line. The centroid gives a frequency of 857.815 MHz, which
corresponds to a redshift of $z_{\rm abs} = 0.65584\pm0.00003$,
cf. the values determined for the Lyman-{$\alpha$} absorption: 0.6561
\citep{rt00} and 0.6564 \citep{clwb98}. The r.m.s. noise is 4.7 mJy
per each 10 \kms\ channel. Bottom: The absorption of Mg{\sc \,i}
(unsaturated) and Mg{\sc \,ii} (saturated) at $z=0.656$ towards
1622+238 (from ESO service-mode program 69.A-0371).}
\label{1622}
\end{figure}

The data were reduced using the {\sc miriad} interferometry reduction
package, with which we extracted a summed spectrum from the emission
region of the continuum map. Upon this, we
found absorption towards the peak of the radio continuum apparent in
both polarisations. After subtracting the continuum with a first order
polynomial from the line free channels of the combined polarisations,
a single Gaussian gives the fit shown in Fig.~\ref{1622}.

The obvious caveat with the detection is the lack of flat baseline in
the spectrum above $\approx300$ \kms\ due to the lost channels.
However, the spectra obtained in each individual polarisation are
fitted by very similar Gaussians. Furthermore, if instead we use
3C\,286 for the bandpass calibration, which was observed at the start
of the run, rather than 3C\,48, which was observed at the end, the
data give very similar profiles (again with the 1 Jy dip in the flux
below 856.6 MHz), which are fitted by the same Gaussian parameters as
before. As would be expected from a real feature towards an unresolved
($<75''\times42''$) background source, the line is present in the
centre and surrounding pixels, but disappears with increasing
proximity from the centre pixel. We also produced a cube in which
1622+238 is offset from the phase centre: While the absorption
was still apparent towards 1622+238, no feature was seen at the 
phase centre. As a further check, we used 3C\,48 to
bandpass calibrate 3C\,286, and, as seen from Fig. \ref{1622} (top
panel), there is no overall ripple in the spectrum.

\section{Results and Discussion}

\subsection{Observational results}
\label{res}

At $\tau=0.0088$, the peak optical depth is at the lower end of the
DLAs detected in 21-cm absorption (table 1 of \citealt{cmp+03}),
although the large line-width gives a velocity integrated optical
depth of $\int\tau dv = 2.2\pm0.2$ \kms, which takes it to joint 5th
place in the line strength stakes (Table \ref{res}).  In the optically
thin regime ($\tau\lapp0.3$), the total atomic hydrogen column
density, $N_{\rm HI}$ [\scm], of the absorbing gas in a homogeneous
cloud is related to the velocity integrated optical depth of the 21-cm
line via
\begin{equation}
N_{\rm HI}=1.823\times10^{18}\,\frac{T_{\rm s}}{f}\int\!\tau\,dv\,,
\end{equation}
where $T_{\rm s}$ [K] is the spin temperature of the gas, $f$ is the
 covering factor of the flux by the absorber and $\int\tau dv$ is the
 velocity integrated optical depth of the 21-cm absorption. If the
 21-cm and Lyman-$\alpha$ absorption arise in the same cloud
 complexes, the total neutral hydrogen column density of
 $2.3\times10^{20}$ \scm\ \citep{rt00},  gives $T_{\rm
 s}/f\approx60$~K, the lowest value yet detected in a
 redshifted source (Table \ref{res}).
\begin{table}
\centering
\begin{minipage}{90mm}
\caption{The DLAs detected in 21-cm absorption listed in order of
increasing spin temperature/covering factor ratio. $z_{\rm abs}$ is
the redshift of the DLA, $\int\tau dv$ is the velocity integrated
optical depth of 21-cm absorption [\kms], $N_{\rm HI}$ is the total
neutral hydrogen column density [\scm] of the absorber (see
\citealt{cwbc01}), which is unknown for the
DLAs towards 0248+430 and 2351+456. The optical identification of the
absorber (ID) is given: D--dwarf, L--LSB, S--spiral, U--unknown.
$T_{\rm s}/f$ is the inferred spin temperature/covering factor ratio
[K]. The final column gives the 21-cm absorption reference. See
\citet{cmp+03} for details. \label{res}}
\begin{tabular}{@{}l c c c c r c@{}} 
\hline
QSO &$z_{\rm abs}$ & $\int\tau dv$& $\log N_{\rm HI}$ & ID & $T_{\rm s}/f$ & Ref. \\
\hline
1622+238$^a$ & 0.6561 & 2.2 & 20.4 & -- & 60 & C07\\
0809+483$^{b}$ & 0.4369 & 0.89 & 20.2 & S & 120 & BM83\\
0235+164 &  0.523385 & 14 & 21.7 & S &200 & R76\\
1629+120 & 0.5318 & 0.49 &  20.5 & S &310 & KC03\\
0458--020 & 2.03945 & 7.2 & 21.7 & U & 380 & W85\\
0827+243 & 0.5247 & 0.26 & 20.3 & S & 420 & KC01\\
1229--021 & 0.39498 & 0.66 & 20.8 & S & 520 & BS79\\
1127--145 & 0.3127 & 2.7 & 21.7 & L & 1000 & L98,CK00\\
1328+307$^c$ & 0.692154 & 0.93 & 21.3 & L & 1200 & BR73\\
0738+313 & 0.2212 & 0.34 & 20.8 & D & 1300 & L98\\
0438--436 & 2.347 & 0.22 & 20.8 & U & 1600 &K06\\
1157+014 & 1.94362 & 2.2 & 21.8 & L& 1600 & W81 \\
0201+113 & 3.386 & 0.71 & 21.3 & U & 2000 & K07\\
0952+179 & 0.2378  & 0.11 & 21.3 & L & 9900 &  KC01\\
\hline
0248+430 & 0.394 & 3.0 & -- & U & -- & LB01 \\
2351+456 & 0.779452 & 13 &-- & U & -- & D04\\
\hline
\end{tabular}
{Notes: $^a$3C\,336, $^{b}$3C\,196, $^{c}$3C\,286.\\ References: BR73:
\citet{br73}, R76: \citet{rbb+76}, BS79: \citet{bs79}, W81: \citet{wbj81}; BM83:
\citet{bm83}, W85: \citet{wbt+85}, L98: \citet{lsb+98}, CK00:
\citet{ck00}, KC01: \citet{kc01a}, LB01: \citet{lb01}, KC03:
\citet{kc02}, D04: \citet{dgh+04}, K06: \citet{kse+06}, K07:
\citet{kcl06}, C07: This paper.}
\end{minipage}
\end{table}
Since both the spin temperature and covering factor are generally unknown, the
 degeneracy in the ratio is best left intact \citep{cmp+03,cw06},
 although for $f\leq1$ (by definition) this means that $T_{\rm
 s}\leq60$ K.

The 21-cm absorption profile in this absorber is much wider than that
found in other DLAs ($\leq53$ \kms, see table 1 of \citealt{cmp+03})
and more like the typical \HI\ emission profile of a galaxy
(e.g. \citealt{ksk+04}). We note also, in the other DLAs the 21-cm
profile-width is always narrower than the metal-ion lines (typically
FWHM$_{\rm 21cm}\approx0.2\,\Delta V_{\rm MgII}$, \citealt{ctp+07}),
but in this case FWHM$_{\rm 21cm}\approx2\,\Delta V_{\rm MgII}$, which
suggests very different kinematics from those of the typical DLA.  In
fact, for other DLAs the spin temperature is often compared with the
kinetic temperature (e.g. \citealt{lb01}), which should be equivalent
for single cloud in thermodynamic equilibrium. Applying $T_{\rm
k}\approx22\times{\rm FWHM}^2$ to 1622+238 gives the unphysical
$T_{\rm k}\sim10^{6}$ K, confirming that kinematical motions are
chiefly responsible for the wide profile.

\subsection{The nature of the {\boldmath$z=0.656$} absorber towards 1622+238}

\label{nat}
\subsubsection{Possible absorber hosts}
\label{poss}

\begin{figure*}
\vspace{15.6cm}
\includegraphics{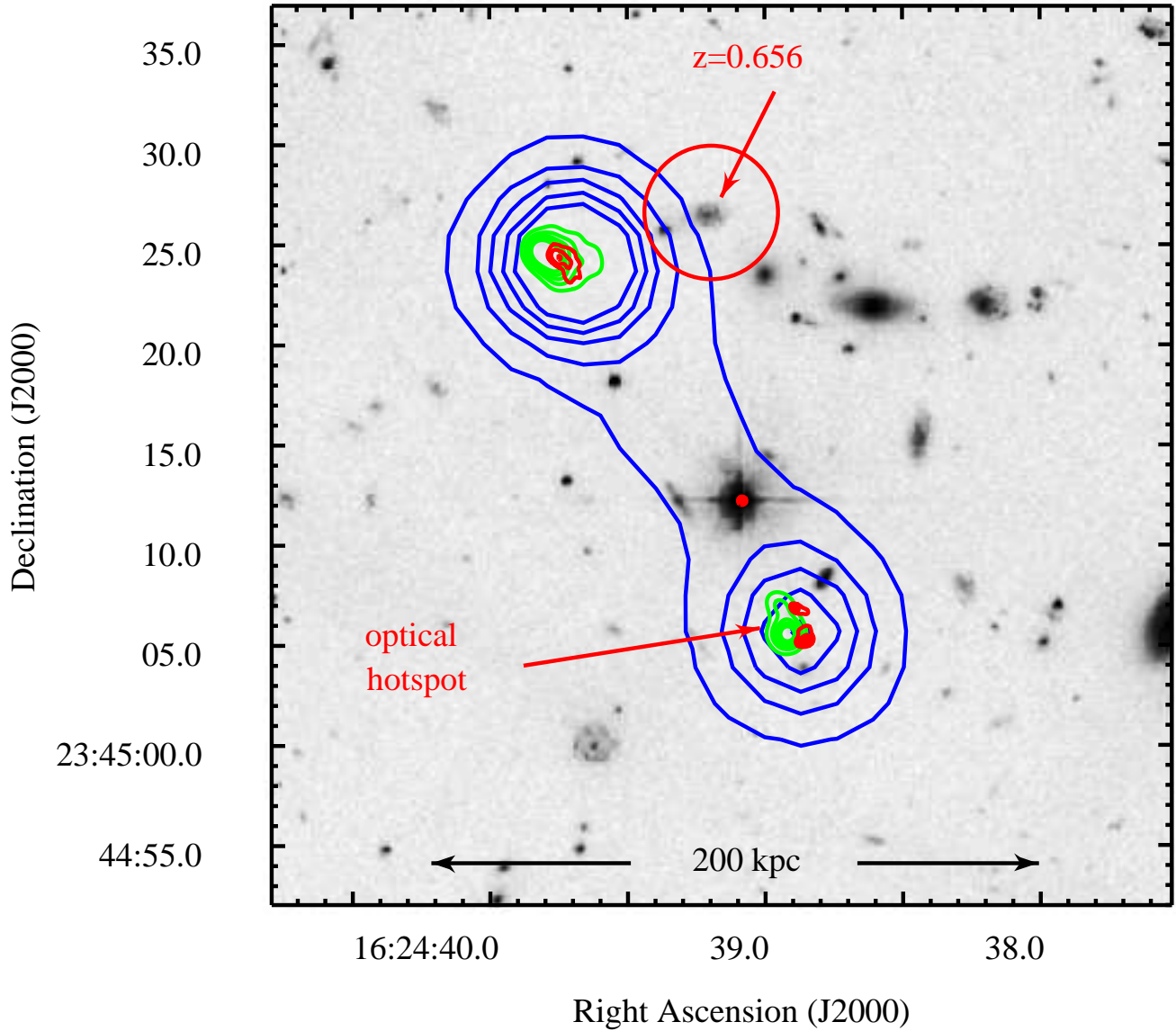}
\caption{The central part of the HST image of \citet{sdm+97} overlaid
with the 8.4 GHz [inner/red contours] and 1.4 GHz [middle/green] radio
emission \citep{cwc05}.  We also show the FIRST 1.4 GHz radio map
[outer lobe/blue]. We do not show the 858 MHz continuum emission from
our observations as the quasar is unresolved by the $75''\times42''$
WSRT beam. The contours scale from 10\% of the peak emission, and for
the sake of clarity, shown only up to 50\% in the 1.4 GHz maps. The
only $z=0.656$ galaxy found by \citet{sdm+97} is highlighted. The
circular region, centered at this galaxy, shows the size of a region
which contains 0.88\% of the total 858 MHz flux density (not
necessarily located here, see text). The linear distance indicated is
calculated for a redshift of $z=0.656$ using
$H_{0}=75$~km~s$^{-1}$~Mpc$^{-1}$, $\Omega_{\rm matter}=0.27$ and
$\Omega_{\Lambda}=0.73$ (used throughout the paper). North is up and
east to the left. The HST image and radio maps are courtesy of Teddy
Cheung and the Very Large Array's FIRST (Faint Images of the Radio Sky
at Twenty Centimetres) survey.}
\label{overlay} 
\end{figure*}

In Fig. \ref{overlay} we show the HST WFPC2 (F702W filter) optical
image of the field towards 3C\,336 ($z_{\rm em}=0.927$) overlaid with
the radio maps. Although there is compact 8.4 GHz emission from the
QSO itself, most of the radio emission is from the extended lobes,
which span an enormous $28''$ \citep{bhl+94,cwc05}.
The DLA towards this QSO is somewhat of a mystery, since the only
galaxy in the HST field with a redshift of $z_{\rm abs}=0.656$ must
have a projected disk impact parameter of $\approx65\,h^{-1}$ kpc
\citep{sdm+97}. Since no other galaxy within this impact parameter has
ever been found, \citet{sdm+97} suggest, but stipulate as unlikely,
that the absorption may be due to an under-luminous late-type spiral,
in which an invisible disk spans at least all of the way from the
identified $z=0.656$ galaxy to the QSO sight-line.

As an alternative, the authors suggest that there may be other
galaxies within this impact parameter, although as these are
undetected, they would have to be extremely faint ($L<0.05L_{\rm
K}^*$). Another possibility is offered by the galaxy of unknown
redshift close to the optical hotspot, or even the hotspot
itself, both of which are located close to a radio lobe
sight-line \citep{cwc05}. If these (and perhaps some other low
luminosity galaxies), were part of a group, which included the
$z=0.656$ galaxy, the redshift criterion would be satisfied, while
perhaps providing a large \HI\ reservoir located closer to the QSO
sight-line. Furthermore, the \HI\ could be largely dispersed over the
region through the mutual tidal interactions of these galaxies
\citep{sdm+97}.

\subsubsection{Implications of the 21-cm absorption}

From Fig. \ref{overlay}, we see that, although the identified
$z=0.656$ galaxy is located close to the sight-line of a radio lobe,
it remains sufficiently remote to be covering a negligible amount of
flux\footnote{Although we would expect the 1420 MHz emission in the
rest frame of the galaxy (858 MHz) to be further extended than the
21-cm emission measured in our rest frame.}. Although we cannot
determine the location of the absorbing region within the radio-bright
region, we may obtain an estimate of its extent: The circular region
centered on the $z=0.656$ galaxy represents the physical area which
encompasses 0.88\% of the total 858 MHz flux. That is, this is the
smallest region that can give the observed optical depth. So although
the visible feature identified as the $z=0.656$ galaxy does not
intercept much radio flux, a cold, invisible disk of $\gapp20$~kpc
(3.3\arcsec ) radius could cover significant radio emission (as shown
in the figure), although, as per the reservations of \citet{sdm+97},
it may not necessarily be centered upon this galaxy. In fact, the
21-cm and Lyman-\AL\ absorption may not be coincident with one another,
in light of the relative profile widths and offset in the profiles
(Fig. \ref{1622}).

For a zero intrinsic velocity offset, one can estimate the variability of the
constant $x\equiv \alpha^2 g_{\rm p} \mu $, where $\alpha$ is the fine
structure constant, $g_{\rm p}$ the proton $g$-factor and $\mu\equiv
m_{\rm e}/m_{\rm p}$ is the electron-to-proton mass ratio. In such a
case, the relative change in the value of $x$ between redshift $z$ and
a terrestrial ($z=0$) value is given by (e.g. \citealt{tw80})
\begin{equation}\label{equ:dxx}
\frac{\Delta x}{x} \equiv \frac{x_z-x_0}{x_0}
= \frac{z_{\rm UV}-z_{\rm 21} } {1+z_{\rm 21}},
\end{equation}
where $z_{\rm 21}$ and $z_{\rm UV}$ represent the observed absorption
redshifts for 21-cm and a UV heavy-element transition,
respectively. Using the Gaussian fit centroid, $z_{\rm
21}=0.65584\pm0.00003$, and the stongest Mg {\sc \,i} component,
$z_{\rm UV}=0.656062\pm0.000003$, we obtain $\Delta x/x =
(1.34\pm0.18)\times 10^{-4}$. This is an order of magnitude larger
than any of the eleven other values which have been determined (see
figure 1 of \citealt{tmw+06}), suggesting that the velocity offset
between the 21-cm and UV absorption is not primarily due to the values
of the fundamental constants at $z=0.656$.

\begin{figure}
\vspace{6.5cm}
\includegraphics{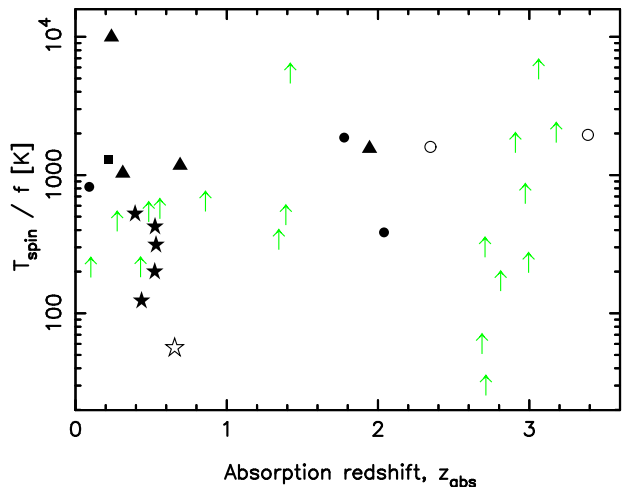}
\caption{The spin temperature/covering factor ratio versus the
absorption redshift for the DLAs searched for in 21-cm absorption. The
symbols represent the 21-cm detections and the shapes represent the
type of galaxy with which the DLA is associated: circle--unknown type,
star--spiral, square--dwarf, triangle--LSB. The arrows show the lower
limits and all of these bar one (0454+039 at $z_{\rm abs}=0.8596$)
have unknown host identifications. The unfilled symbols show the new
detections since \citet{cmp+03}, from which the figure is adapted:
\citet{kse+06,kcl06} and 1622+238 (unfilled star), on the assumption
that the 21-cm absorption arises in the same gas as the Ly-\AL\
absorption.}
\label{Toverf}
\end{figure}

Therefore the observed $\approx50$ km s$^ {-1}$ velocity difference,
in conjunction with the atypical 21-cm line-width, suggests a physical
offset between the cold, 21-cm absorbing, and warm, UV absorbing,
gas. However, although not completely coincident, hydrogen is observed
in both regimes, within the same field (possibly remote from the QSO
sight-line), at (nearly) the same redshift, and so we find it unlikely
that the radio and UV absorption are not related in some way. The
width of the 21-cm absorption in comparison to the optical lines, and
that of other DLAs, may be the result of an atypical situation
where we have an extended absorber occulting an extended radio
source. However, all of the known DLAs associated with spiral galaxies
also occult large radio sources (figure 4 of
\citealt{cmp+03}). Furthermore, although located much closer to the
spirals than the other DLAs in terms of $T_{\rm s}/f$, due to the wide
profile, the 21-cm absorption towards 1622+238 may be somewhat of an
anomaly (Fig.~\ref{Toverf})\footnote{Note that if FWHM$_{\rm
21cm}\approx0.2\,\Delta V_{\rm MgII}$, as per the other DLAs, $T_{\rm
s}/f$ would be $\approx600$ K, placing this bang in the middle of the
spirals in Fig.~\ref{Toverf}.}.

If the system is bound and the wide 21-cm profile is purely the result
of rotational broadening (i.e. not an outflow), the total line-width
of $\pm280$ \kms, leads to a dynamical mass estimate of $M_{\rm
dyn}=1.8\times10^{10}\,r/\sin^2 i$ \Mo, where $r$ [kpc] is the edge of
the cold \HI\ distribution [or the 858 MHz emission, whichever comes
first]. In the case where this gas is located close to the QSO sight-line, so
that the identified $z=0.656$ galaxy is actually near the edge of
the absorbing disk or group\footnote{Assuming $i$ is the inclination
of the most rapidly orbiting galaxies and neglecting their individual
rotations.}, i.e. $r\approx75$ kpc \citep{clwb98}, the dynamical mass
becomes $M_{\rm dyn}\sim1\times10^{12}/\sin^2 i$ \Mo. If the gas is
due to a large spiral, centered on the $z=0.656$ galaxy\footnote{Which
may be indicated by the single Gaussian fit, although signal-to-noise
ratio of the spectrum (Fig. \ref{1622}) is far too low to resolve
possible contributions from smaller discrete galaxies.}, by assuming
circular motions and the same inclination as the visible ``nucleus''
(55\dg, \citealt{clwb98}), the deprojected velocities and impact
parameter give $M_{\rm dyn}\sim2\times10^{12}$ \Mo\ (for $r\approx75$
kpc). Although large, the derived masses are close to the values
derived for the two most local spirals, the Milky Way ($M_{\rm
dyn}\sim2\times10^{12}$~\Mo, \citealt{we99}) and M31 ($M_{\rm
dyn}\sim1\times10^{12}$~\Mo, \citealt{ewg+00}), and certainly
reasonable for a group of smaller galaxies.

Combining this dynamical mass with the luminosity of $L<0.25\,L^*$
\citep{sdm+97}\footnote{Note that, from local studies, the vast
majority (87\%) of DLAs are believed to be due to galaxies of
luminosities $<L^*$ \citep{zvb+05}.}, gives a mass-to-light ratio of
$M_{\rm dyn}/L\gapp400$ \Mo/\Lo. Such ratios ($\sim1000$ \Mo/\Lo),
define the hypothesised low density galaxies (LDGs), which can range
in mass from $10^{9}$ to $10^{12}$\Mo\ \citep{jhhp97}. In an LDG a low
density disk resides in a dark halo, with the surface density below
that required by the Toomre criterion for instabilities to form
($\lapp10^{20}$ \scm, \citealt{jhhp97,voj02}), and so there is no
ongoing star formation: From deep HST H\AL\ imaging, \citet{blc+01}
determine a star formation rate of $<0.15$ \Mo\ yr$^{-1}$ kpc$^{-2}$
in the 3C\,336 field. Another possible example of an LDG may already
have been observed as the ``dark hydrogen cloud'', VIRGOHI 21 in the
Virgo Cluster \citep{mdd+05}. This has a mass-to-light ratio of
$M_{\rm dyn}/L_B> 500$ \Mo/\Lo\ and spans at least 16 kpc in
extent. If it is a galaxy, this has a mass of $M_{\rm dyn}>10^{11}$
\Mo, although, as may be the case towards 1622+238, the large amount
of gas could be due to tidal debris. Where ever the \HI\ (Ly-\AL\ \&
21-cm) absorption arises towards 1622+238, it is clear that this would
not be the first time a large amount of unidentified neutral hydrogen
has been detected.

\section{Summary}

We have detected 21-cm absorption in the $z_{\rm abs}=0.656$ damped
Lyman-$\alpha$ absorber towards 1622+238 (3C\,336). If the absorption
arises in the same gas as the Lyman-$\alpha$ absorption, the spin
temperature of the gas is $T_{\rm s}\leq60$ K, the lowest yet found in
a DLA. The low spin temperature is a consequence of the extremely wide
profile, which, most atypically, is wider than the Mg{\sc \,ii} line.
This could be due to a large absorber occulting what we know to be
large radio source. The 21-cm is also offset from the unsaturated
Mg{\sc \,i} absorption by $\approx-50$ \kms, which corresponds to
$\Delta x/x \approx1.3\times 10^{-4}$ ($x\equiv \alpha^2 g_{\rm p} \mu
$), which is an order of magnitude larger than that expected from
other DLAs, suggesting that the offset is not dominated by different
values of the fundamental constants at $z=0.656$. It may indicate,
instead, that the radio and UV absorption are unrelated, although a
lower Lyman-$\alpha$ column density would give gas significantly
colder than 60 K. Furthermore, it seems somewhat strange that the
hydrogen absorbers detected in two different regimes at the same
redshift would be unrelated, especially considering that none of these
seems to arise directly along the QSO sight-line.
Regarding this, our results cannot distinguish from where the hydrogen
absorption arises, whether this be the disk of a large underluminous
spiral, an extremely faint galaxy lying close to the QSO sight-line or
a possible tidal interaction, between other faint galaxies, which is
dragging significant amounts of gas across the sight-line.

\section*{Acknowledgments}



We would like to thank Raffaella Morganti for coordinating the WSRT
observations, Alan Bridle and Teddy Cheung for their radio images, and
Bob Carswell, Martin Zwaan, Steve Longmore, Alan Pedlar, Rob Beswick,
Matt Owers and Matthew Whiting for their advice. This research has
made use of the NASA/IPAC Extragalactic Database (NED) which is
operated by the Jet Propulsion Laboratory, California Institute of
Technology, under contract with the National Aeronautics and Space
Administration.  This research has also made use of NASA's
Astrophysics Data System Bibliographic Services.


\label{lastpage}
\end{document}